\documentclass{webofc}

\usepackage[varg]{txfonts}   
\usepackage{hyperref}
\usepackage{url}
\hypersetup{colorlinks=true,citecolor=blue,urlcolor=blue,linkcolor=blue}
\usepackage{multirow}
\usepackage{braket}

\begin{document}
\title{Understanding photon TMDs using light-front wave function}
%
%

\author{\firstname{Satyajit} \lastname{Puhan}\inst{1}\fnsep\thanks{\email{puhansatyajit@gmail.com}} \and
        \firstname{Narinder} \lastname{Kumar}\inst{1,2}\fnsep\thanks{\email{narinderhep@gmail.com}} \and
        \firstname{Harleen} \lastname{Dahiya}\inst{1}\fnsep\thanks{\email{dahiyah@nitj.ac.in}}
}

\institute{Computational High Energy Physics Lab, Department of Physics, Dr. B.R. Ambedkar National Institute of Technology, Jalandhar, 144008, India  
\and
Computational Theoretical High Energy Physics Lab, Department of Physics, Doaba College, Jalandhar 144004, India. }

\abstract{In this work, we try to understand the transverse structure of photon through transverse momentum dependent parton distribution functions (TMDs) in quantum chromodynamics. We calculate all the possible time reversal photon TMDs using light-front wave functions. For this work, we have considered photon as a Fock-state of quark anti-quark pair. All the $9$
 T-even TMDs have been presented in the overlap and explicit form of light-front wave functions using the helicity amplitudes. We observe that only $3$
 TMDs are non-zero for the case of a real photon as compared to $8$
 for virtual photon. We briefly discuss the unpolarized $f_1
 (x,\textbf{k}_\perp)$, longitudinally polarized $g_{1L}(x,\textbf{k}^2_\perp)$, transversly polarized $h_1(x,\textbf{k}^2_\perp)$ and tensor polarized $f_{1LL}(x,\textbf{k}^2_\perp)$ TMDs for both real and virtual photon.}
\maketitle
\section{Introduction}
\label{intro}
Understanding the hadron structure using the multi-dimensional generalized parton distribution functions (GPDs), transverse momentum parton distribution functions (TMDs) \cite{Diehl:2015uka}, generalized parton distribution functions (GTMDs) \cite{Lorce:2011dv}, parton distribution functions (PDFs) etc. are very crucial and challenging. It is even more difficult to understand the photon internal structure compared to proton, pion, kaon and other hadrons in quantum field theory and hadron physics. Photon being a gauge boson of quantum electrodynamics (QED), does not have a static internal structure in a way as a hadron but because of its quantum fluctuations into quark anti-quark pairs, it can behave as a hadronic particle at high energies. 
To understand the photon internal structure, one has to probe the high energy photon instead of hadron in the deeply virtual Compton scattering (DVCS) process as $\gamma^* (Q)+ \gamma \rightarrow \gamma +\gamma$. There have been many theoretical studies to understand the photon distribution amplitudes (DAs) \cite{Ball:2002ps}, GPDs \cite{Nair:2023lir}and TMDs. OPAL \cite{OPAL:1999rcd}, L3 \cite{L3:1998ijt}, PLUTO \cite{PLUTO:1984gmq}, CELLO \cite{CELLO:1983crq} and DELPHI \cite{Delphi} Collaboration have provided a wide range of data for the $F^\gamma_2$ structure function of photon in QED. In this work, we mainly target the transverse structure of real and virtual photon through their possible TMDs, while treating the photon as a Fock-state of quark anti-quark pair. For numerical predictions, we have used the light-front wave functions in the overlap form of helicity amplitudes. As we are treating the photon as a spin-$1$ meson, there are total $9$ T-even TMDs present at the leading twist \cite{Puhan:2023hio}. However in this work, we have discussed only the unpolarized $f_1(x,\textbf{k}^2_\perp)$, longitudinal polarized $g_{1L}(x,\textbf{k}^2_\perp)$, transversely polarized $h_1(x,\textbf{k}^2_\perp)$ and tensor polarized $f_{1LL}(x,\textbf{k}^2_\perp)$ TMDs. For a better understanding transverse structure,  we have also predicted the TMDs not only for the real photon but also for the virtual photon 

\section{Methodology}

\subsection{Light-front formalism for photon}\label{LFQMe}
In QED, photon can be expressed in terms of an electron positron pair along with a bare photon as \cite{Nair:2023lir}
\begin{eqnarray}
	| \gamma \rangle = | \gamma \rangle_{bare} \Psi _{\gamma}  + |e^- e^+ \rangle \Psi _{e^- e^+}.
\end{eqnarray}
However, in the high energy regime of QCD, photon can fluctuate into quark anti-quark pair. In this way, one can express the photon Fock-state similar to mesons as \cite{Xiao:2003wf,Pasquini:2024qxn}
\begin{eqnarray}
	|\gamma\rangle = \sum   | q \bar q \rangle \Psi _{q \bar q}+.......
	\label{eigen}
\end{eqnarray}
In this work, we have taken Eq. (\ref{eigen}) as the eigenstate of the photon to treat photon like a spin-$1$ meson. The two-particle Fock state representation for the photon with different helicities ($\Lambda= \pm 1, 0$) and all possibilities of helicities of quark and anti-quark \cite{Xiao:2003wf}
\begin{eqnarray}
	|\Psi_{\gamma} (P^+, P_{\perp},h_{q},h_{\bar q})\rangle &=& \int \frac{dx d^2\textbf{k}_{\perp}}{16 \pi^3} [\Psi^{\Lambda} (x,\textbf{k}_{\perp},\uparrow,\downarrow)|(x P^+,\textbf{k}_{\perp},\uparrow,\downarrow)\rangle\nonumber \\
	&& + \Psi^{\Lambda} (x,\textbf{k}_{\perp},\uparrow,\uparrow)|(x P^+,\textbf{k}_{\perp},\uparrow,\uparrow)\rangle \nonumber \\
	&& +\Psi^{\Lambda} (x,\textbf{k}_{\perp},\downarrow,\uparrow)|(x P^+,\textbf{k}_{\perp},\downarrow,\uparrow)\rangle\nonumber \\
	&&+ \Psi^{\Lambda} (x,\textbf{k}_{\perp},\downarrow,\downarrow)|(x P^+,\textbf{k}_{\perp},\downarrow,\downarrow)\rangle].
	\label{photon}
	\end {eqnarray}
	Here, $P=(P^+, P^-,{\bf P_{\perp}})$ and $k=(k^+, k^-,k_{\perp})$ are the four vector momentum of the photon and active quark respectively. $x={k^+}/{P^+}$ is the longitudinal momentum fraction carried by the quark, $h_{q(\bar q)}$ are the different helicity possibilities of quark (anti-quark) respectively. In the above equation, $\Psi^{\Lambda} (x,\textbf{k}_{\perp},h_{q},h_{\bar q})$ is the dynamic spin wave functions and can be expressed as \cite{Xiao:2003wf}
	\begin{eqnarray}
		\Psi^{\Lambda} (x,\textbf{k}_{\perp},h_{q},h_{\bar q}) &=& \chi^{\Lambda} (x,\textbf{k}_{\perp},h_{q},h_{\bar q}) \phi_{\gamma}(x, \textbf{k}_{\perp}),
		\label{wave}
	\end{eqnarray}
	where $\phi_{\gamma}(x, \textbf{k}_{\perp})$ is the momentum space wave function for the photon can be expressed as \cite{Xiao:2003wf,Brodsky:2000ii}
	\begin{eqnarray}
		\phi_{\gamma}(x, \textbf{k}_{\perp})=\frac{e_{q}}{M_{\gamma}-(\textbf{k}^2_{\perp}+m^2)/x -(\textbf{k}^2_{\perp}+m^2)/(1-x)}.
		\label{mome}
	\end{eqnarray}
    Here, $m$ and $M_\gamma$ are the quark mass and photon mass respectively. In this work, we have taken quark mass $m=0.2$ GeV. For the real photon, $M_{\gamma}=0$, whereas $M^2_{\gamma}>0$ and $M^2_{\gamma}<0$ for the case of time-like and space like virtual photon respectively \cite{Nair:2023lir}.
$\chi^{\Lambda} (x,\textbf{k}_{\perp},h_{q},h_{\bar q})$ in Eq. (\ref{wave}) is the Lorentz invariant spin structure of the photon expressed  by accounting the photon-quark-anti-quark vertex as \cite{Kaur:2020emh,Xiao:2003wf,Qian:2008px}
	\begin{eqnarray}
        \chi^{\Lambda} (x,\textbf{k}_{\perp},h_{q},h_{\bar q}) = \bar{u}(k_1, h_q) \frac{1}{\sqrt{2} M} 
\left( \frac{k_1 - k_2}{M + 2m}-\gamma \right). 
\epsilon_{\Lambda}(P)  v(k_2, h_{\bar q}).
    \end{eqnarray}
	Here, $k_1$ and $k_2$ denote the four momenta of quark and anti-quark respectively. $(1-x)$ is the momentum fraction carried by the anti-quark from photon and $\epsilon_\Lambda$ is the polarization vector of the photon. $M$ can be defined as $M^2=(m^2+k^2)/x(1-x)$. Now using the longitudinal and transverse polarization of photon, the spin wave function $\chi^{\Lambda} (x,\textbf{k}_{\perp},h_{q},h_{\bar q})$ can be written as for $\Lambda=T(+1)$
    
	\begin{equation}
\left\{
  \begin{array}{lll}
    \chi^{T(+1)} (x,\textbf{k}_{\perp},\uparrow,\uparrow)&=&w^{-1}[\mathbf{k}_\perp^2+(M+2 m)m]\\
    \chi^{T(+1)} (x,\textbf{k}_{\perp},\uparrow,\downarrow)&=&w^{-1}[k^R(xM+m)]\\
    \chi^{T(+1)} (x,\textbf{k}_{\perp},\downarrow,\uparrow)&=&w^{-1}[-k^R((1-x)M+m)]\\
    \chi^{T(+1)} (x,\textbf{k}_{\perp},\downarrow,\downarrow)&=&w^{-1}[-(k^R)^2].
  \end{array}
\right.
\end{equation}
Similarly, for longitudinal photon polarization $\Lambda=0$
\begin{equation}
\left\{
  \begin{array}{lll}
    \chi^{L} (x,\textbf{k}_{\perp},\uparrow,\uparrow)&=&\frac{1}{\sqrt{2}}w^{-1}[k^L((1-2x)M]\\
    \chi^{L} (x,\textbf{k}_{\perp},\uparrow,\downarrow)
            &=& \frac{1}{\sqrt{2}}w^{-1}[2 \mathbf{k}_\perp^2 + (M+2 m)m]\\
     \chi^{L} (x,\textbf{k}_{\perp},\downarrow,\uparrow)
            &=& \frac{1}{\sqrt{2}}w^{-1}[2 \mathbf{k}_\perp^2 + (M+2 m)m]\\
     \chi^{L} (x,\textbf{k}_{\perp},\downarrow,\downarrow)&=&\frac{1}{\sqrt{2}}w^{-1}[-k^R((1-2x)M].
\end{array}
\right.
\end{equation}
 For $\Lambda=T(-1)$, we have
\begin{equation}
\left\{
  \begin{array}{lll}
     \chi^{T(-1)} (x,\textbf{k}_{\perp},\uparrow,\uparrow)&=&w^{-1}[-(k^L)^2]\\
    \chi^{T(-1)} (x,\textbf{k}_{\perp},\uparrow,\downarrow)&=&w^{-1}[k^L((1-x)M+m)]\\
    \chi^{T(-1)} (x,\textbf{k}_{\perp},\downarrow,\uparrow)&=&w^{-1}[-k^L(xM+m)]\\
    \chi^{T(-1)} (x,\textbf{k}_{\perp},\downarrow,\downarrow)&=&w^{-1}[\mathbf{k}_\perp^2+(M+ 2 m)m].
\end{array}
\right.
\end{equation}
	It is important to mention here that the above spin wave functions for the polarized photon satisfy the spin sum rule  $\Lambda=h_{q}+h_{\bar q}+L_{z}$. For the case of photon, different configurations of
	the light-front wave function with orbital angular momentum (OAM) $L_z = 0$, $\pm1$ and $\pm2$ correspond to the S, P and D wave components respectively.

    \begin{figure}[ht]
		\centering
		\begin{minipage}[c]{1\textwidth}\begin{center}
				(a)\includegraphics[width=.45\textwidth]{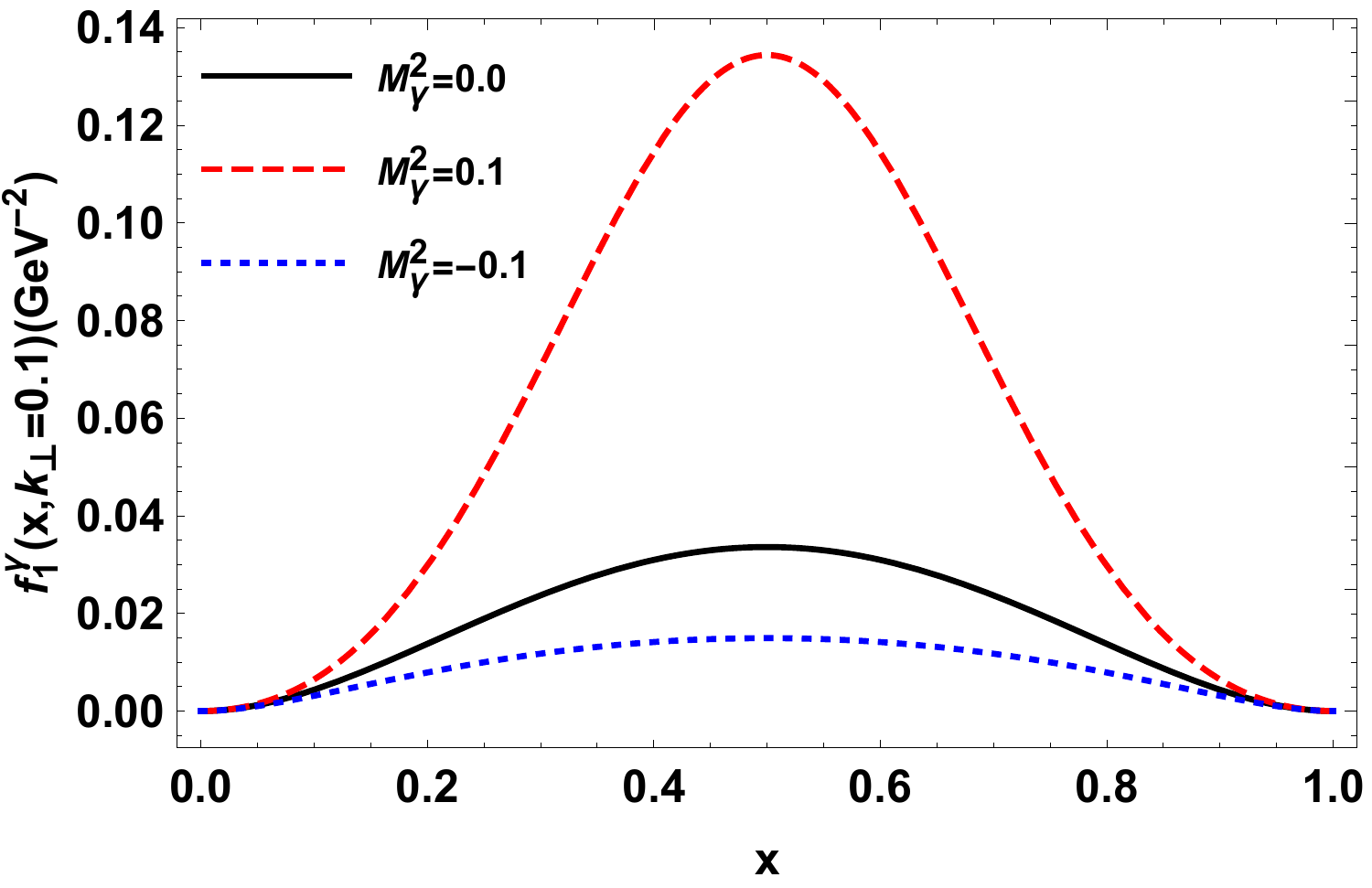}
				(b)\includegraphics[width=.45\textwidth]{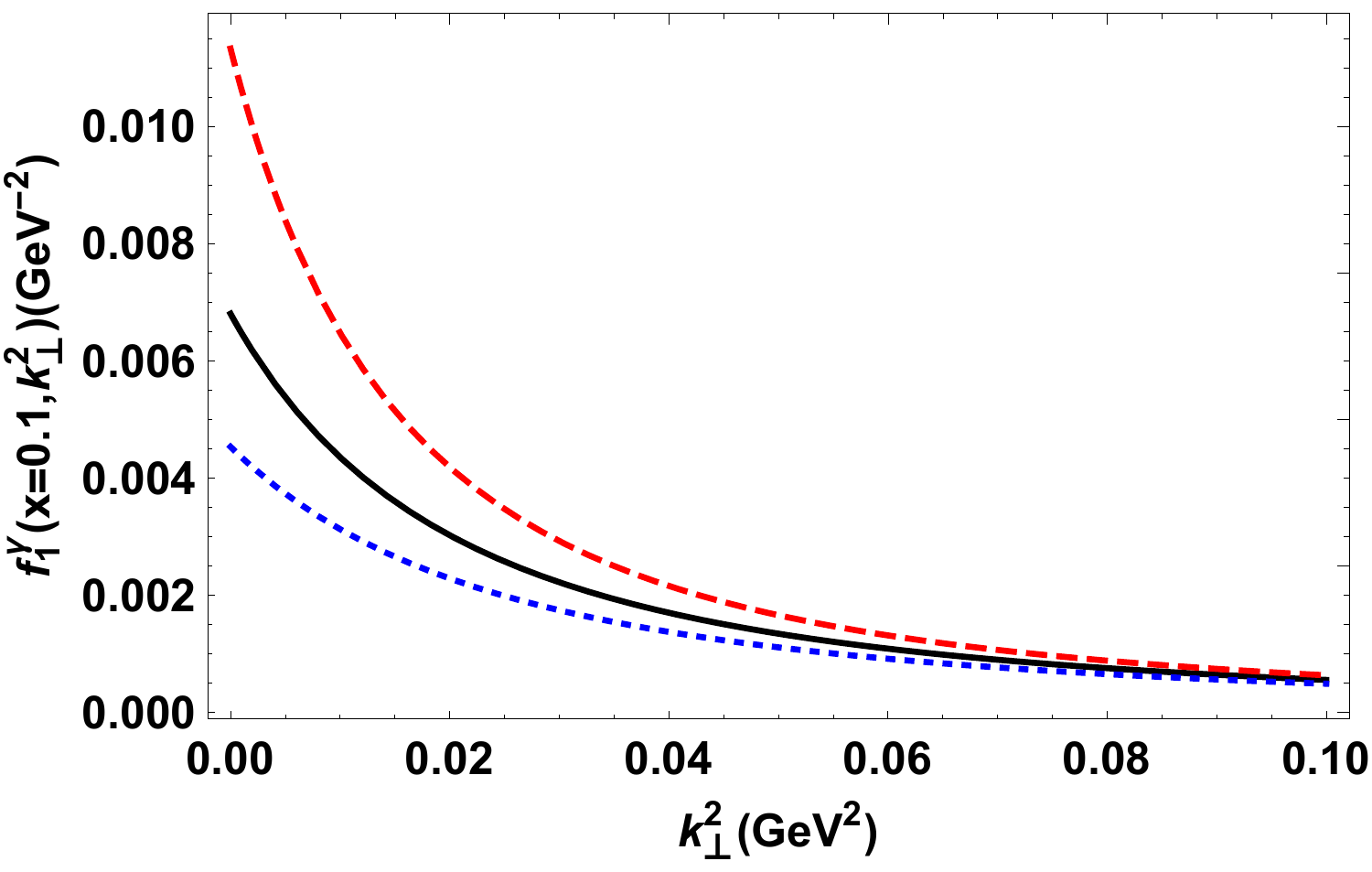}
			\end{center}
		\end{minipage}
		\begin{minipage}[c]{1\textwidth}\begin{center}
				(c)\includegraphics[width=.45\textwidth]{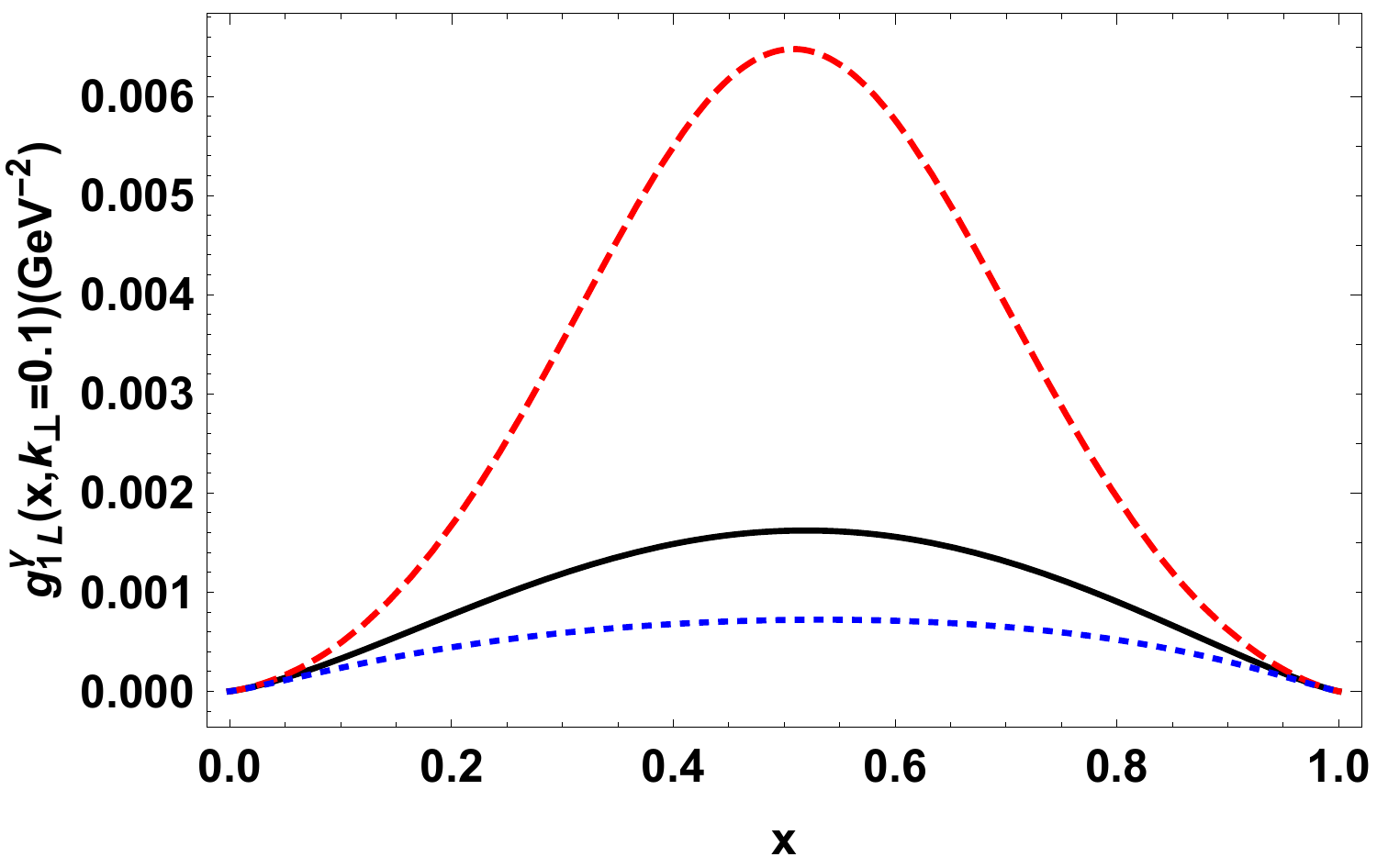}
				(d)\includegraphics[width=.45\textwidth]{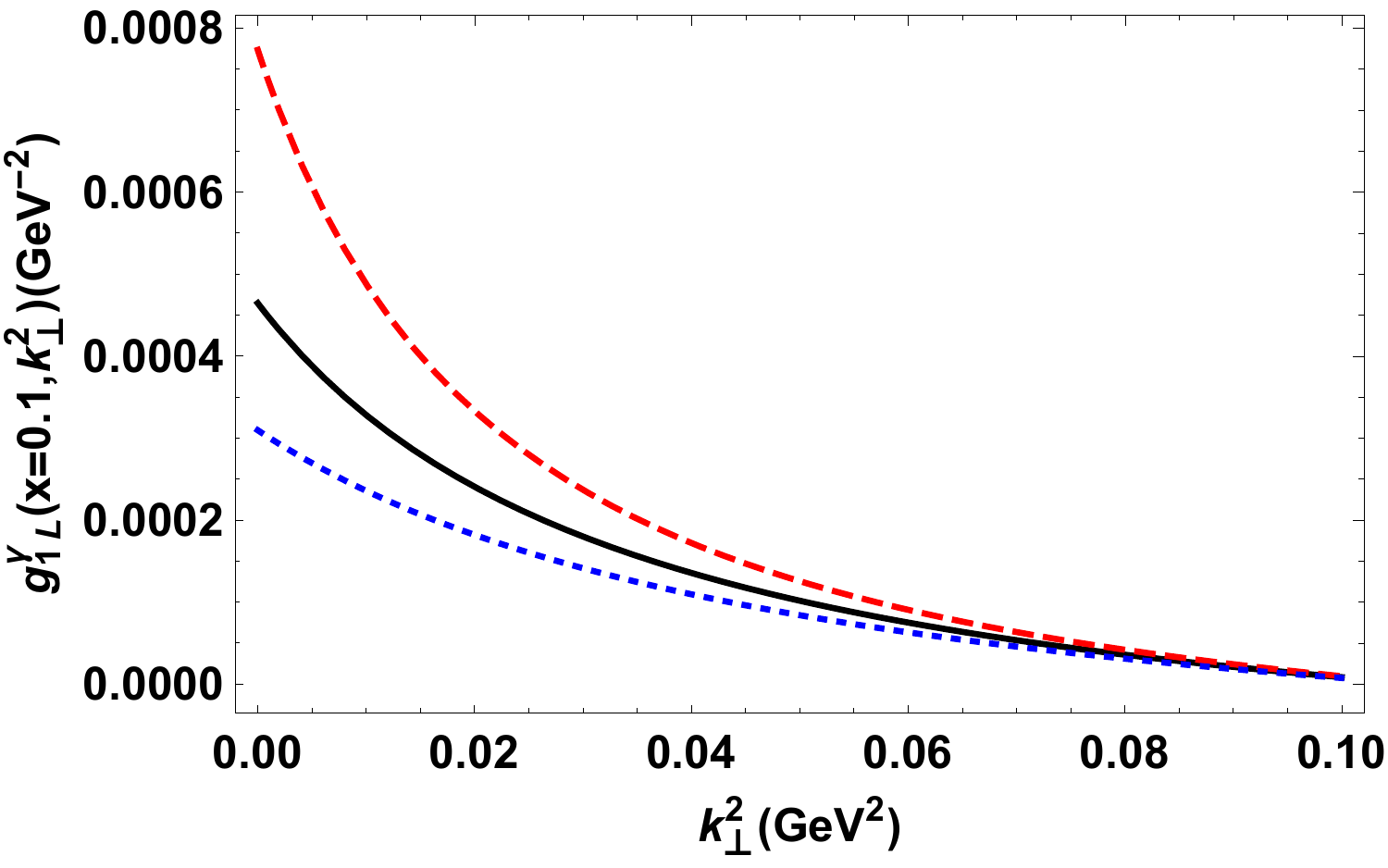}
                \end{center}
		\end{minipage}
        \begin{minipage}[c]{1\textwidth}\begin{center}
				(e)\includegraphics[width=.45\textwidth]{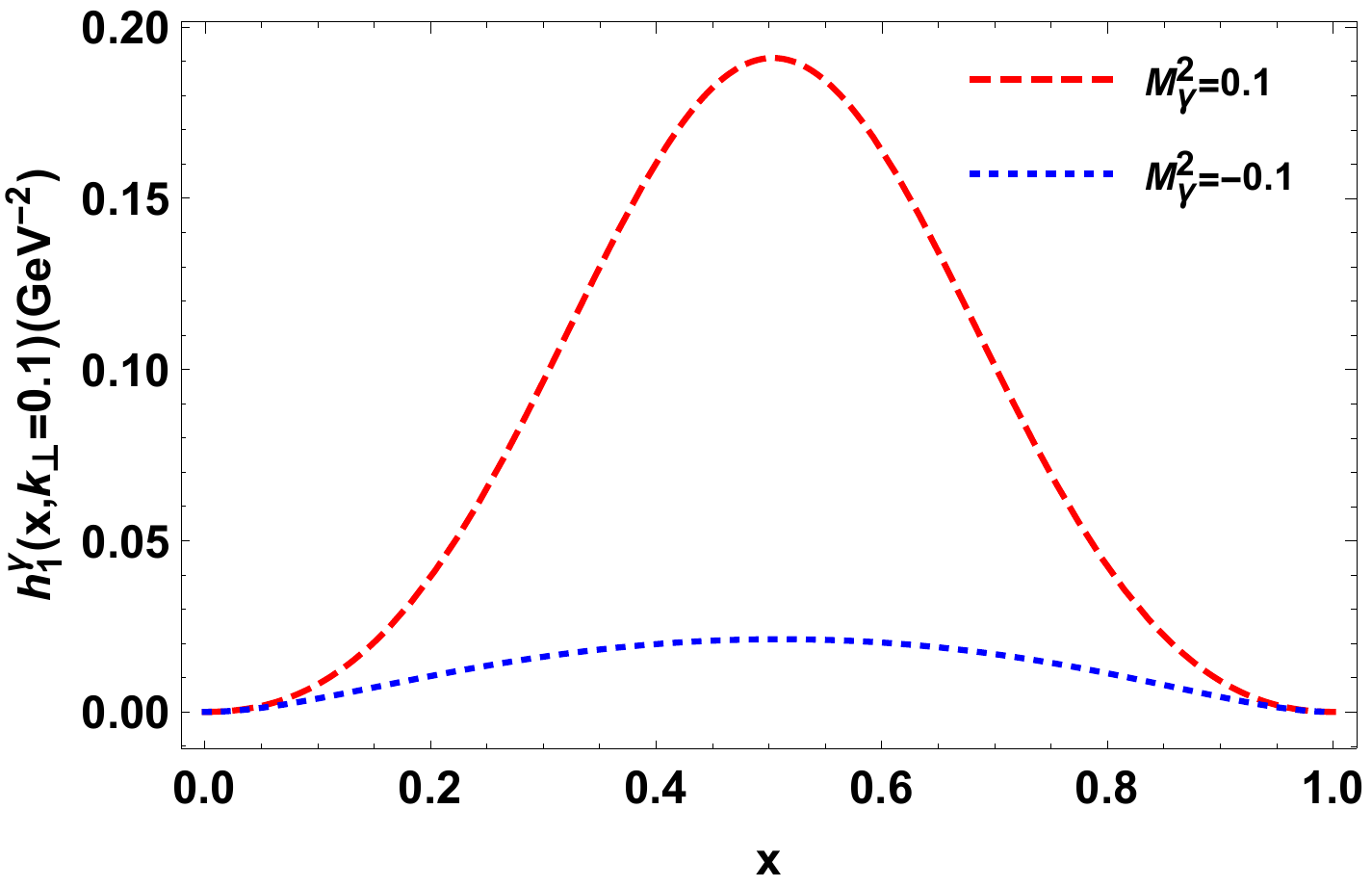}
				(f)\includegraphics[width=.45\textwidth]{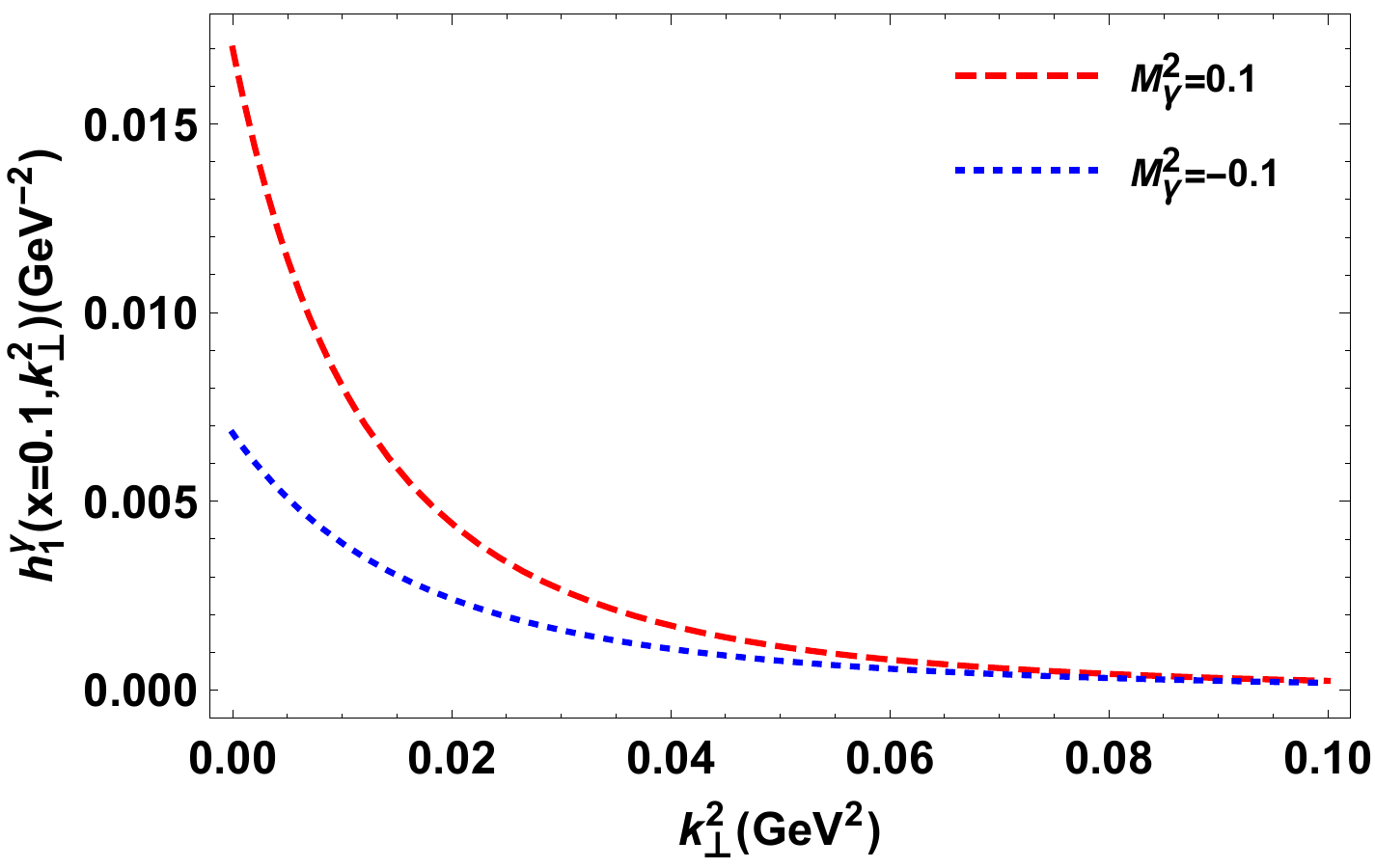}
			\end{center}
		\end{minipage}
        \begin{minipage}[c]{1\textwidth}\begin{center}
				(g)\includegraphics[width=.45\textwidth]{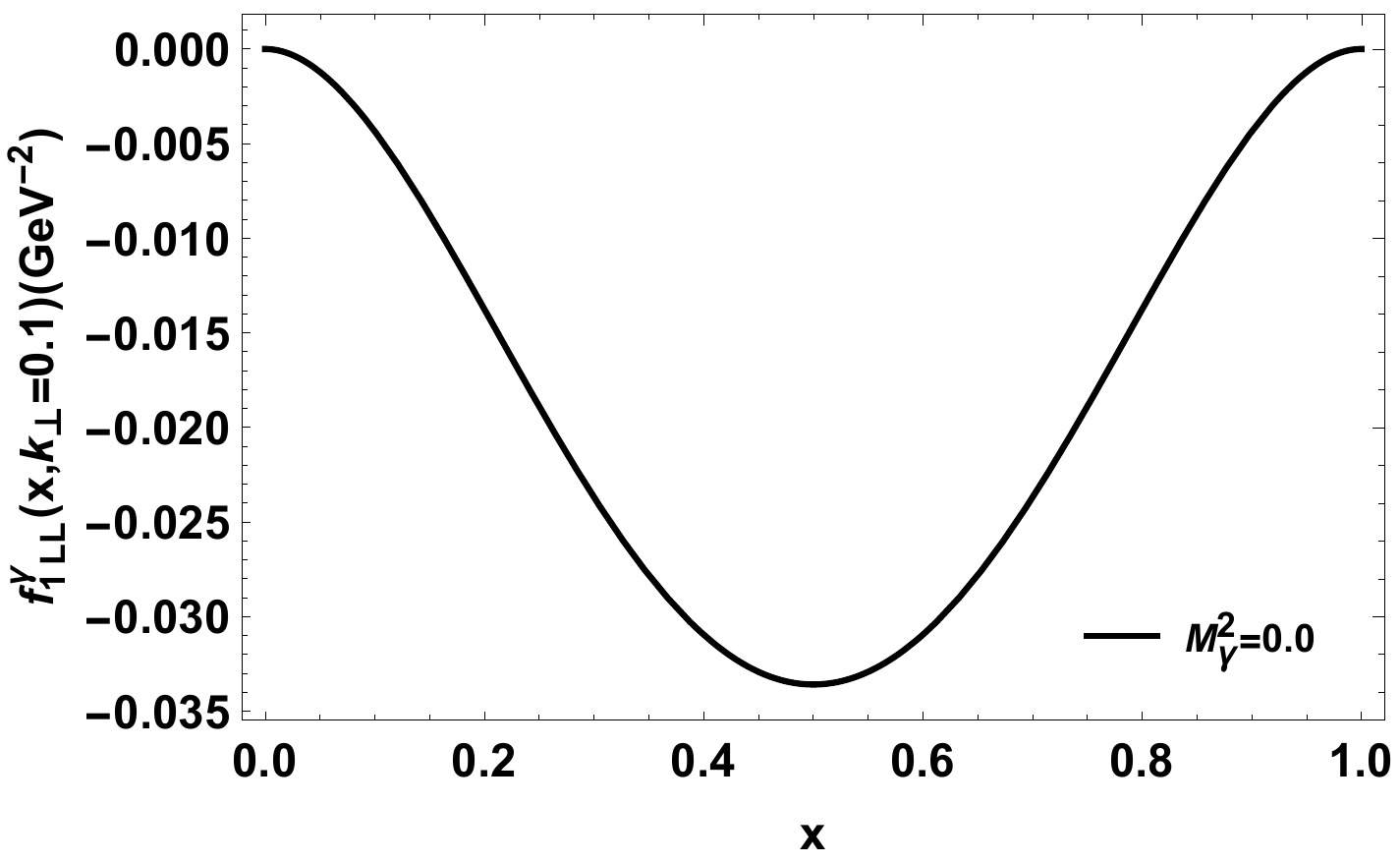}
				(h)\includegraphics[width=.45\textwidth]{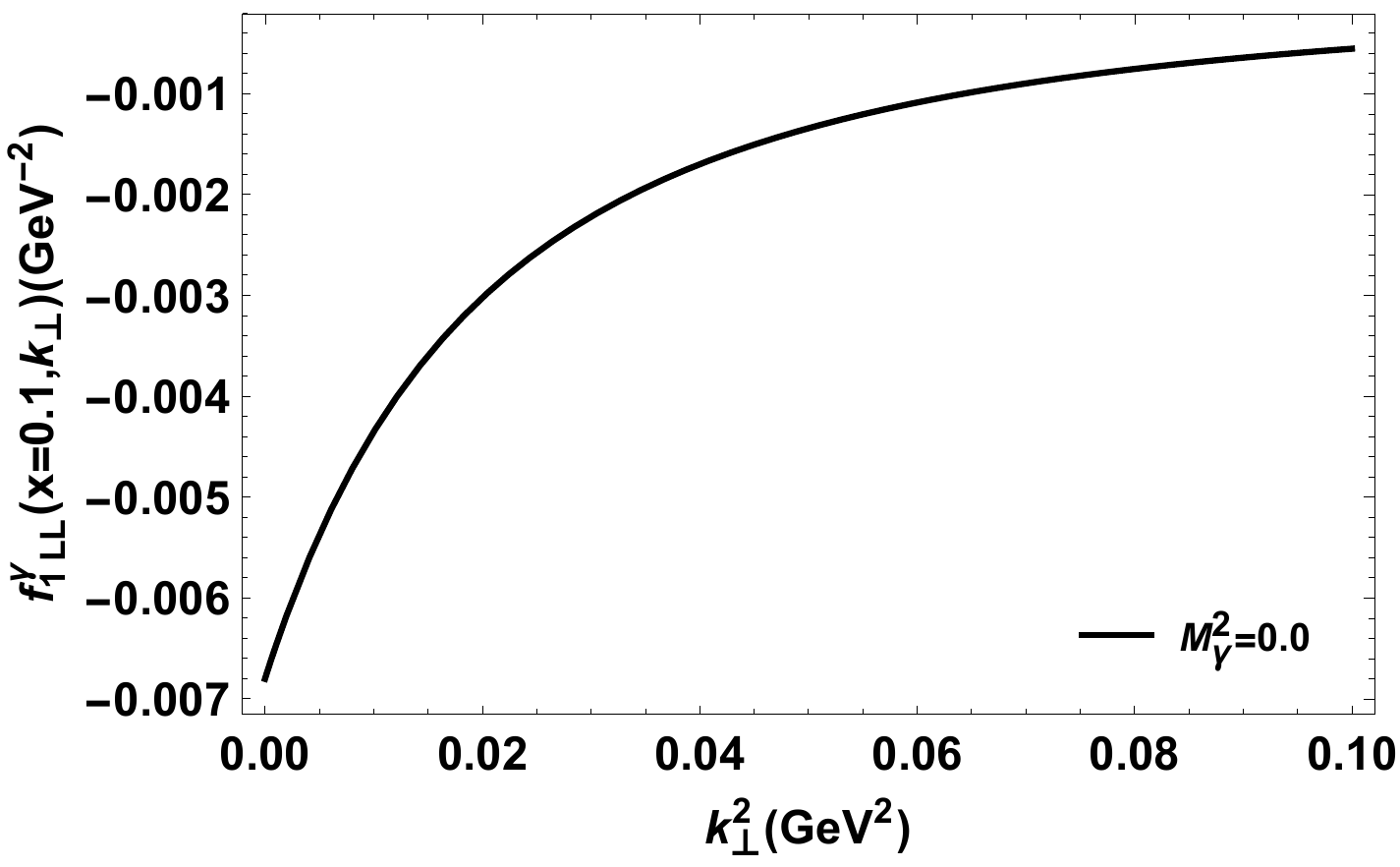}
			\end{center}
		\end{minipage}
		\caption{(Color online) The TMDs for real photon ($M^2_\gamma=0$ GeV$^2$), time-like virtual photon ($M^2_\gamma=0.1$ GeV$^2$), and space-like virtual photon ($M^2_\gamma=-0.1$ GeV$^2$) have been plotted with respect to longitudinal momentum fraction (x) at fixed value transverse momenta $\textbf{k}_\perp=0.1$ GeV in the left panel and with respect to $\textbf{k}^2_\perp$ GeV$^2$ at fixed $x=0.1$ in the right panel.}
		\label{realtmds}
	\end{figure}

\subsection{Transverse momentum parton distribution functions}
There are total nine possible T-even TMDs at the leading twist of spin-1 target, which are expressed as unpolarized ($U$), transversely polarized ($T$) and longitudinally polarized ($L$) quark and hadron polarizations. There are three extra tensor polarized TMDs present for the case of spin-1 targets than the spin-$1/2$ nucleons. In this work, we have only discussed $f_1(x,\textbf{k}^2_\perp)$, $g_{1L}(x,\textbf{k}^2_\perp)$,  $h_1(x,\textbf{k}^2_\perp)$ and $f_{1LL}(x,\textbf{k}^2_\perp)$ TMDs only. 
These four TMDs can be expressed in the form helicity amplitude as \cite{Shi:2022erw,Puhan:2023hio}
\begin{eqnarray*}
		f_{1}(x,\mathbf{k}_{\perp}^2) &=& \frac{1}{6}(\mathcal{A}_{\uparrow L,\uparrow L}+\mathcal{A}_{\downarrow L,\downarrow L}+\mathcal{A}_{\uparrow T(+),\uparrow T(+)}+\mathcal{A}_{\downarrow T(+),\downarrow T(+)} \nonumber\\
		&& +\mathcal{A}_{\uparrow T(-),\uparrow T(-)}+ \mathcal{A}_{\downarrow T(-),\downarrow T(-)}), \\
		g_{1L}(x,\mathbf{k}_{\perp}^2) &=& \frac{1}{4}(\mathcal{A}_{\uparrow T(+),\uparrow T(+)}-\mathcal{A}_{\downarrow T(+),\downarrow T(+)}-\mathcal{A}_{\uparrow T(-),\uparrow T(-)}+\mathcal{A}_{\downarrow T(-),\downarrow T(-)}), \\
        h_{1}(x,\mathbf{k}_{\perp}^2) &=& \frac{1}{4\sqrt{2}}(\mathcal{A}_{\uparrow T(+),\downarrow L}+\mathcal{A}_{\downarrow L,\uparrow T(+)}+\mathcal{A}_{\uparrow L,\downarrow T(-)}+\mathcal{A}_{\downarrow T(-),\uparrow L}), \\
        f_{1LL}(x,\mathbf{k}_{\perp}^2) &=& \frac{1}{2}(\mathcal{A}_{\uparrow L,\uparrow L}+\mathcal{A}_{\downarrow L,\downarrow L})-\frac{1}{4}\left(\mathcal{A}_{\uparrow T(+),\uparrow T(+)}+\mathcal{A}_{\downarrow T(+),\downarrow T(+)}+\mathcal{A}_{\uparrow T(-),\uparrow T(-)}\right. \nonumber \\
		&&\left. +\mathcal{A}_{\downarrow T(-),\downarrow T(-)}\right), \\
        \end{eqnarray*}
where the helicity amplitude can be expressed in the overlap form of light-front wave function as      

\begin{eqnarray}
		\mathcal{A}_{h^\prime_q \Lambda^\prime, h_q \Lambda}(x,{\bf k}^2_\perp)&=&\frac{1}{(2 \pi)^3} \sum_{h_q, h_{\bar{q}}} \Psi^{\Lambda^\prime *}(x,{\bf k}_\perp,h^\prime_q, h_{\bar{q}})\,\Psi^{\Lambda}(x,{\bf k}_\perp,h_q, h_{\bar{q}})\, .
		\label{O1}
	\end{eqnarray}

	The OAM along z-axis is conserved for all the TMDs. The functions $f_1$, $g_{1L}$, and $f_{1LL}$ have zero OAM transfer from initial state to final state photon. The TMDs $f_1$, $g_{1L}$and $h_1$ describe the momentum distributions of the unpolarized
quark in the unpolarized photon, the longitudinally polarized quark in the longitudinally
polarized photon, and the transversely polarized quark in the transversely polarized
photon, respectively. 

\section{Results} 
For the numerical predictions, we have taken the quark mass $m=0.2$ GeV and real photon mass $M_\gamma=0$ GeV. We have also taken into account both space-like and time like virtual photons by taking $M^2_\gamma=0.1$ GeV$^2$ and $M^2_\gamma=-0.1$ GeV$^2$ respectively. For the case of real photon, we have only transverse mode polarizations as the longitudinal mode is zero. Therefore, the helicity amplitudes having longitudinal polarization of LFWFs are zero. We have plotted unpolarized $f_1(x,\textbf{k}^2_\perp)$, longitudinal polarized $g_{1L}(x,\textbf{k}^2_\perp)$, transversely polarized $h_1(x,\textbf{k}^2_\perp)$ and tensor polarized $f_{1LL}(x,\textbf{k}^2_\perp)$ TMDs for both the real and virtual photons at fixed $x=0.1$ and $\textbf{k}_\perp^2=0.1$ GeV$^2$ in Fig. \ref{realtmds}. We have observed that the TMDs show symmetry around $x=0.5$ and vanish after $\textbf{k}^2_\perp \ge 0.1$ GeV$^2$. Except $f_{1LL}(x,\textbf{k}^2_\perp)$ TMD, other TMDs show positive distribution for both the real and virtual photon. The time-like virtual photon shows higher peak values compared to real photon and space-like virtual photon. For the case of real photon, the $h_1(x,\textbf{k}^2_\perp)$ TMD comes out to be zero and similarly $f_{1LL}(x,\textbf{k}^2_\perp)$ TMD is zero for the case of virtual photon. As the longitudinal mode of real photon is zero, the $f_1(x,\textbf{k}^2_\perp)$ and $f_{1LL}(x,\textbf{k}^2_\perp)$ TMDs have equal and opposite distributions as seen from Fig. \ref{realtmds} (a) and (g). 

This work is very essential to study the fundamental structure of photon. In addition, exclusive vector meson generation in virtual photon-proton or photon-nucleus
scattering can be studied using the virtual photon LFWFs.

\section*{Acknowledgement}
 H.D. would like to thank the Science and Engineering Research Board, Anusandhan-National Research Foundation, Government of India under the scheme SERB-POWER Fellowship (Ref No. SPF/2023/000116) for financial support.  N.K. would like to thank the Science and Engineering Research Board, Anusandhan-National Research Foundation, Government of India under the scheme SERB-TARE (Ref No. TAR/2021/000157) for financial support.

\end{document}